\documentclass[11pt,fleqn]{article}

\usepackage{amsmath,amssymb}
\usepackage{graphicx}
\usepackage[top=0.75in, bottom=0.75in, left=0.75in, right=0.75in]{geometry}
\usepackage[small]{caption}
\usepackage[noblocks]{authblk}
\usepackage{hyperref}
\usepackage{cite}

\usepackage[usenames,dvipsnames]{color}

\begin{document}

\title{\sf  Next-to-Leading Order (NLO) Perturbative Effects in QCD Sum-Rule Analyses of Light Tetraquark Systems: A Case Study in the Scalar-Isoscalar Channel
}
\author[1,2]{B.A. Cid-Mora\thanks{bac302@usask.ca}}
\author[1]{T.G. Steele\thanks{tom.steele@usask.ca}}

\affil[1]{Department of Physics \&
Engineering Physics, University of Saskatchewan, SK, S7N 5E2, Canada}
\affil[2]{GSI Helmholtzzentrum f\"{u}r Schwerionenforschung, Darmstadt, Germany}

\maketitle

\begin{abstract}
QCD sum-rule mass predictions for tetraquark states provide insights on the interpretations and internal structure of experimentally-observed exotic mesons. However, the overwhelming majority of tetraquark QCD sum-rule analyses have been performed at leading order (LO), which raises questions about the underlying theoretical uncertainties from higher-loop corrections. The impact of next-to-leading order (NLO) perturbative effects are systematically examined in scalar ($J^{PC}=0^{++}$) isoscalar light-quark tetraquark systems where comprehensive LO sum-rule analyses have been performed and NLO perturbative corrections to the correlators have previously been calculated. Using the scalar-isoscalar  state as a detailed case study to illustrate the differences between LO and NLO analyses, it is shown that NLO effects in individual Laplace sum-rules are numerically significant and have an important role in improving the reliability of the sum-rule analyses by widening the Borel window. However, ratios of sum-rules are found to be less sensitive to NLO effects with the additional advantage of cancelling the anomalous dimension that emerges from the NLO corrections. NLO mass predictions based on these sum-rule ratios are thus remarkably robust despite the slow perturbative convergence of the underlying correlator. The mass predictions $0.52\,\text{GeV}< m_{\sigma}< 0.77\,\text{GeV}$ 
for the lightest scalar-isoscalar $\sigma$  state are in good agreement with the four-quark interpretation of the $f_0(500)$, and the relative coupling strengths of the $f_{0}(980)$ and $f_0(500)$ to the tetraquark current agree with the pattern found in chiral Lagrangian analyses. Effects of the $\sigma$ resonance width are studied for  different models, including resonance shapes inspired by chiral Lagrangians.
\end{abstract}

\section{Introduction}
The 2003 discovery of the $X(3872)$ by the Belle collaboration \cite{Belle:2003nnu} marked the beginning of a new era in hadronic physics, and there are now a multitude of observed mesons that do not fit the patterns of the quark-antiquark ($q\bar q$) conventional quark model. Four-quark ($qq\bar q\bar q$) states in a variety of flavour and colour (e.g., tetraquark, meson molecules) configurations provide a compelling framework for these exotic mesons (see {e.g.,} Refs.~\cite{Liu:2019zoy,Brambilla:2019esw,Swanson:2006st,Godfrey:2008nc} for reviews). Indeed, four-quark interpretations seem particularly inevitable for (charmonium-like) charged states such as the $Z_c(4430)$ and $Z_c(3900)$ \cite{Belle:2007hrb,BESIII:2013ris,Belle:2013yex}.  Furthermore, in light quark systems one expects an inverted mass hierarchy (see e.g., Ref.~\cite{Jaffe:1976ig}) which provides a tetraquark framework for the lightest scalar mesons.

QCD sum-rules are based on quark-hadron duality and parameterize non-perturbative effects via QCD condensates, permitting the prediction of resonance properties through QCD correlation functions of composite operators that serve as hadronic interpolating fields \cite{Shifman:1978bx,Shifman:1978by}. There is a vast literature associated with QCD sum-rule applications to hadronic systems (see e.g., Refs.~\cite{Reinders:1984sr,narison,Gubler:2018ctz}), with specific applications to four-quark systems reviewed in Refs.~\cite{Liu:2019zoy,Albuquerque:2018jkn} (also note the early work on light-quark molecular systems in Ref.~\cite{Latorre:1985uy}). The overwhelming majority of these QCD sum-rule analyses of four-quark systems are at leading-order (LO) in the perturbative contributions, which raises obvious questions about the effect of higher-loop corrections on the mass predictions. By contrast, sum-rule analyses of conventional $q\bar q$ mesons contain at least next-to-leading order (NLO) perturbative contributions, even in the seminal studies of Refs.~\cite{Shifman:1978bx,Shifman:1978by}.

There are some exceptions to the general trend of LO sum-rule analyses of four-quark systems. Within the factorization approximation, NLO effects have been considered for four-quark systems containing heavy quarks, and these estimates indicate reasonable control of higher-order perturbative effects on the mass predictions \cite{Albuquerque:2016znh,Albuquerque:2017vfq,Albuquerque:2020hio,Albuquerque:2020ugi,Albuquerque:2021tqd} (see Refs.~\cite{Narison:1994zt,Hagiwara:2002hf,Lucha:2019pmp} for different findings on non-factorizable contributions). However, in light-quark systems exact calculations of NLO perturbative effects have been performed for scalar-isoscalar $\sigma$ currents with tetraquark ({\it i.e.,} diquark-antidiquark) composition \cite{Groote:2014pva}. For the optimized $\sigma$ currents identified in the comprehensive LO sum-rule analysis \cite{Chen:2007xr}, the NLO effects are surprisingly large \cite{Groote:2014pva} indicating slow perturbative convergence of the QCD correlation function, motivating a detailed study comparing the LO and NLO QCD sum-rule analyses and mass estimates for these systems.

In this paper, the scalar-isoscalar channel for light tetraquark currents is used as a detailed case study to compare the LO and NLO QCD sum-rule analyses and mass predictions. In particular, the full QCD Laplace sum-rule methodology (including establishing the Borel window, parameter optimization, and mass predictions) is compared for the LO and NLO QCD correlation functions to explore the influence of higher-loop perturbative effects.

In Section \ref{LSR_sec} the QCD currents and correlation functions for the scalar-isoscalar $\sigma$ channel are presented along with a summary of QCD Laplace sum-rule methodology. In Section~\ref{methodology_section} the role of NLO effects is examined within individual sum-rules and in ratios with an emphasis on the differences in Borel windows resulting from LO and NLO effects, motivating a focus on ratios which mitigate the large NLO effects and anomalous dimensions. The NLO sum-rule analysis based on these sum-rule ratios is presented in Section~\ref{prediction_section} for a variety of resonance models. Results are summarized and discussed  in Section \ref{conclusions_section}.

\section{NLO Contributions to Laplace QCD Sum-Rules}
\label{LSR_sec}
QCD sum rules probe hadronic properties through correlation functions $\Pi\left(Q^2\right)$ of composite operators $J(x)$ reflecting the quark/gluonic valence content of the state
\begin{equation}
\Pi\left(Q^2\right)= i\int d^4x \,e^{i q\cdot x}\left\langle \Omega  \right\vert  T\left( J(x)J^\dagger(0) \right)\left\vert \Omega \right\rangle\,,~ Q^2=-q^2\,.
\label{corr_fn}
\end{equation}
For the dimension-six four-quark operators of interest, the QCD correlation function is connected to the hadronic regime via a dispersion relation
\begin{equation}
\Pi\left(Q^2\right)=\Pi(0)+Q^2\Pi'(0)+\frac{1}{2}Q^4\Pi''(0)
+\frac{1}{3}Q^6\Pi'''(0)
+Q^8
\int\limits_{t_0}^\infty
\,\frac{ \rho(t)}{t^4\left(t+Q^2\right)}dt \,,
\label{GenDispRel}
\end{equation}
where the hadronic spectral function $\rho(t)$ with physical threshold $t_0$ contains states $\left\vert h\right\rangle$ with quantum numbers that can be interpolated to the vacuum through the operator $J$ such that $\left\langle h\right\vert J\left\vert \Omega \right\rangle\ne 0$. Excited states can be suppressed and the unknown subtraction constants can be eliminated in \eqref{GenDispRel} via the Borel transform operator $\hat B$ \cite{Shifman:1978bx,Shifman:1978by} 
\begin{equation}
\hat B\equiv 
\lim_{\stackrel{N,~Q^2\rightarrow \infty}{N/Q^2\equiv \tau}}
\frac{\left(-Q^2\right)^N}{\Gamma(N)}\left(\frac{d}{dQ^2}\right)^N\,, 
\label{BorelOp}
\end{equation}
which has the following properties
\begin{gather}
\hat B\left[a_0+a_1Q^2+\ldots a_m Q^{2m}\right]=0\,,~ m=0,1,2,\ldots \quad (m~{\rm finite})\,,
\label{BorelPoly}\\
 \hat B \left[ \frac{Q^{2n}}{t+Q^2}\right]=\tau \left(-1\right)^nt^n\mathrm{e}^{-t\tau}  \,,~ n=0,1,2,\ldots 
\quad (n~{\rm finite})\, .
\label{BorelExp}
\end{gather}
A family of Laplace sum-rules are now obtained by applying $\hat B$ to \eqref{GenDispRel} weighted by integer powers of $Q^2$, which will involve the QCD (theoretical) entity
\begin{equation}
{\cal L}_k(\tau)\equiv\frac{1}{\tau}\hat B\left[\left(-1\right)^k Q^{2k}\Pi\left(Q^2\right)\right]
\label{laplace}
\end{equation}
connected to hadronic spectral function by
\begin{equation}
{\cal L}_{k}(\tau)=\int\limits_{t_0}^\infty
t^k e^{-t\tau}\rho(t)\,dt \,,~ k= 0,1,2,\ldots \, .
\label{GenLap}
\end{equation}
Although the Laplace sum-rules \eqref{GenLap} exponentially suppress excited states compared to the ground state, in cases where there are mixed interpretations an excited state may have an enhanced coupling that partially compensates for the exponential suppression. 

A comprehensive study of the Laplace sum-rules of the light scalar ($J^{PC}=0^{++}$) states, using a basis of five independent tetraquark currents, identified the following mixed scalar-isoscalar $\sigma$ current $J$ as having the best theoretical properties for a LO Laplace sum-rule analysis \cite{Chen:2007xr}
\begin{gather}
   J = \cos(\theta)J_{A_{6}^{\sigma}}+\sin(\theta)J_{V_{3}^{\sigma}}\,,~ \cot{\theta} = \frac{1}{\sqrt{2}}\,,
    \label{eq:current1}
    \\
   J_{A_{6}^{\sigma}}= (u^{iT}C\gamma^{\mu}d^{j})(\bar{u}_{i}\gamma_{\mu}C\bar{d}_{j}^{T}+\bar{u}_{j}\gamma_{\mu}C\bar{d}_{i}^{T})\,,
   \label{A6_currrent}
   \\
    J_{V_{3}^{\sigma}}= (u^{iT}C\gamma^{\mu}\gamma_{5}d^{j})(\bar{u}_{i}\gamma_{\mu}\gamma_{5}C\bar{d}_{j}^{T}-\bar{u}_{j}\gamma_{\mu}\gamma_{5}C\bar{d}_{i}^{T}) \,,
       \label{V3_currrent}
\end{gather}
where $\{i, j\}$ represent colour degrees of freedom. The notation $A_6$ indicates a combination of axial vector substructures in the six-dimensional colour representation, and $V_3$ indicates vector substructures in the three-dimensional colour representation.  

The QCD prediction for the Laplace sum-rule is represented in terms of the QCD spectral function $\rho^{\rm QCD}$ as 
\begin{equation}
{\cal L}^{{\rm QCD}}_{k}(\tau)=\int\limits_{0}^\infty
s^k e^{-s\tau}\rho^{{\rm QCD}}(s)\,ds\,.
\label{GenQCDLap}
\end{equation}
 The quantity $\rho^{{\rm QCD}}(s)$ can be expressed as perturbative contributions $\rho^{{\rm pert}}(s)$ and QCD condensate terms 
\begin{equation}
    \rho^{{\rm QCD}}(s) = \rho^{{\rm pert}}(s)
    + \Biggl(\frac{6\sqrt{2}+7}{2304\pi^{5}}\langle \alpha_{s}GG\rangle +\frac{(m_{u}+m_{d})\langle \bar{q}q\rangle}{36\pi^4} \Biggr)s^{2} \\ 
   +\frac{6\sqrt{2}+1}{288\pi^{3}}(m_{u}+m_{d})\langle \bar{q}q \rangle\langle \alpha_{s}GG\rangle 
   \,,
\label{eq:rho_QCD}
\end{equation}
where the QCD condensate corrections were calculated in Ref.~\cite{Chen:2007xr} and numerically-small subleading quark mass terms have been omitted. Numerical values for the condensates and other QCD input parameters will be outlined below. The perturbative corrections are expressed as the LO and NLO contributions 
\begin{gather}
\rho^{{\rm pert}}(s)= \rho^{{\rm LO}}(s)+  \rho^{{\rm NLO}}(s)\,,\\
\rho^{{\rm LO}}(s)=\frac{s^{4}}{11520\pi^6}  \,,
\label{rho_LO}
\\
\rho^{{\rm NLO}}(s)=\frac{s^{4} }{11520\pi^6}
\frac{\alpha_{s}(\mu)}{\pi}\Biggl [\frac{409-192\sqrt{2}}{40} +\frac{7-6\sqrt{2}}{4}\log\Bigl(\frac{\mu^{2}}{s} \Bigr)\Biggr]
\label{rho_NLO}
\end{gather}
where the LO contributions were calculated in Ref.~\cite{Chen:2007xr} and the NLO contributions were calculated in the chiral limit for the $\overline{\rm MS}$ scheme with renormalization scale $\mu$ in Ref.~\cite{Groote:2014pva}.\footnote{We are grateful to S.~Groote for clarifying a factor of 2 typo in Eq.~(74) of \cite{Groote:2014pva} which has been corrected in \eqref{rho_NLO}. }

The NLO perturbative contributions in \eqref{rho_NLO} imply that $\rho(s)$ satisfies an inhomogeneous renormalization-group (RG) equation containing an anomalous-dimension term $\gamma_\rho$
\begin{gather}
 0=   \left[ \mu \frac{\partial}{\partial\mu}+
 \beta\left(\alpha_s\right) \alpha_s \frac{\partial}{\partial\alpha_s}-2\gamma_\rho
 \right]\rho(s)
 \\
 \beta\left(\alpha_s\right)=\beta_1\frac{\alpha_s}{\pi}+\ldots\,,~\beta_1=-\frac{11}{2}+\frac{n_f}{3}\,,
 \\
 \gamma_\rho=\gamma_1\frac{\alpha_s}{\pi}+\ldots\,,~\gamma_1=
 \frac{7-6\sqrt{2}}{4}\,,
\label{gamma_rho}
\end{gather}
where the factor $\sqrt{2}$ in $\gamma_1$ ultimately originates from the current mixing angle in \eqref{eq:current1}. In principle, the anomalous dimension term prevents application of the established methodology for RG improvement of Laplace sum-rules for light-quark systems \cite{Narison:1981ts}. However, up to NLO the quantity
\begin{equation}
\tilde\rho(s)=    \left(\alpha_s\right)^{2\gamma_1/\beta_1} \rho(s)\,,
\label{tilde_rho}
\end{equation}
satisfies a homogeneous RG equation so that RG improvement of the Laplace sum-rules can then be achieved via $\mu^2=1/\tau$ \cite{Narison:1981ts}. Alternatively, \eqref{tilde_rho} implies that ratios of Laplace sum-rules, or indeed ratios of different contributions (e.g., perturbative and non-perturbative) will be independent of the anomalous dimension. Where appropriate, our analysis will therefore focus on sum-rule ratios because they will not be affected by potentially large and unknown higher-order contributions to $\gamma_\rho$.  

\begin{figure}[htb]
    \centering
    \includegraphics[scale=1.2]{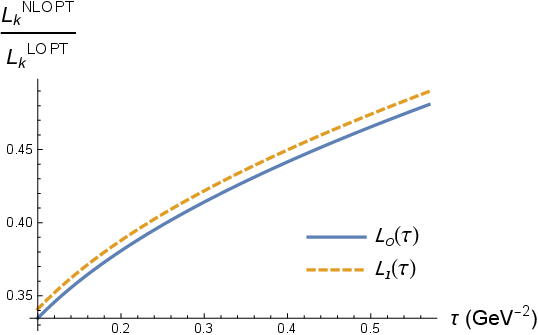}
    \caption{The ratio of NLO and LO perturbative contributions to 
    ${\cal L}^{\rm QCD}_k(\tau)$ is shown as a function of $\tau$ for $k=0$ (solid blue curve) and $k=1$ (dashed orange curve). Note that this ratio is independent of the anomalous dimension factor. } 
    \label{Lk_NLO_fig}
\end{figure}

With a methodology now established for the RG improvement of QCD Laplace sum-rules, the numerical significance of NLO effects can be investigated. The necessary QCD input is the one-loop, $\overline{\rm MS}$ strong coupling for $n_f=4$ active quark flavours appropriate to benchmark values at the tau lepton mass scale \cite{ParticleDataGroup:2020ssz}
\begin{gather}
\alpha_{s}(\mu) = \frac{\alpha_{s}\left(M_{\tau}\right)}%
{1+\frac{25}{12\pi}\alpha_s(M_{\tau})\log\left(\frac{\mu^2}{M_{\tau}^2}\right)}\,,
\label{running_coupling}
\\
  M_\tau = 1.77686 \pm0.00012\,{\rm GeV}\,,~
  \alpha_{s}\left(M_{\tau}\right) = 0.325\pm 0.015\,.
\end{gather}
Fig.~\ref{Lk_NLO_fig} shows that the NLO perturbative contributions are numerically significant compared to the LO effects in the QCD Laplace sum-rules. For example, at a typical $2\,{\rm GeV}$ Borel mass scale corresponding to $\tau=0.25\,{\rm GeV^{-2}}$ the NLO contributions are 40\% of LO, motivating a detailed Laplace sum-rule analysis to examine the effect of these large NLO contributions.

\section{Analysis Methodology for Laplace Sum-Rules}
\label{methodology_section}
QCD Laplace sum-rules are founded on global quark-hadron duality, which equates the hadronic and QCD contributions in \eqref{GenLap}
\begin{equation}
{\cal L}_k^{\rm QCD}(\tau)=\int\limits_{t_0}^\infty
t^k e^{-t\tau}\rho^{\rm had}(t)\,dt \,,~ k= 0,1,2,\ldots \, ,
\label{LSR_had}
\end{equation}
where ${\cal L}^{\rm QCD}(\tau)$ is given via \eqref{GenQCDLap}, and $\rho^{\rm had}(t)$ is the hadronic spectral function with physical threshold $t_0$ appropriate to the quantum numbers associated with the operators (currents) used to construct the correlation function \eqref{corr_fn}. The exponential factor in \eqref{LSR_had} suppresses the high-energy region, thereby enhancing the lowest-lying hadronic resonance(s) in the spectral function. The hadronic spectral function is then split into a resonance contribution and a QCD continuum (see e.g., Refs.~\cite{Shifman:1978bx,Shifman:1978by,Reinders:1984sr,narison,Gubler:2018ctz}) 
\begin{gather}
\rho^{\rm had}(t)=\rho^{\rm res}(t)   + \theta\left(t-s_0 \right) \rho^{\rm QCD}(t)\,,
\label{had_rho}
\\
c_k\left( \tau,s_0\right)=\int\limits_{s_0}^\infty
t^k e^{-t\tau}\rho^{\rm QCD}(t)\,dt\,,
\end{gather}
resulting in the standard form of the Laplace sum-rule relating the QCD prediction ${\cal R}_k\left(\tau,s_0\right)$ to the hadronic resonance contributions
\begin{gather}
{\cal R}_k\left(\tau,s_0\right) ={\cal L}_k^{\rm QCD}(\tau)-c_k\left( \tau,s_0\right)\,,
\\
{\cal R}_k\left(\tau,s_0\right)=\int\limits_{t_0}^{s_0}
t^k e^{-t\tau}\rho^{\rm res}(t)\,dt\,.
\label{LSR_final}
\end{gather}
Although the resonance contributions are integrated to the threshold $s_0$, it is generally expected that there will be a duality gap that separates the resonance and QCD continuum contributions.

The methodology for analyzing the Laplace sum-rules \eqref{LSR_final} consists of three key elements: determining a ``Borel window" on $\tau$ where the analysis is theoretically reliable; selecting a resonance model containing a number of parameters (e.g., hadronic masses); and an optimization method to determine both the resonance parameters and the QCD continuum $s_0$.   

Determining the upper bound on $\tau$ for the Borel window is associated with the non-perturbative (QCD condensate) contributions to ${\cal L}_k^{\rm QCD}(\tau)$ as mass dimension increases in comparison to perturbative contributions. As shown below, the dominant non-perturbative contribution comes from the dimension-four gluon condensate $\langle \alpha_s GG\rangle$ term in \eqref{eq:rho_QCD} and hence a convergence criterion is defined via the quantity
\begin{equation}
    B_k=\frac{{\cal L}^{\langle \alpha GG\rangle}_k(\tau)}{{\cal L}^{\rm pert}_k(\tau)}<\frac{1}{3},
    \label{B_k}
\end{equation}
which requires that the gluon condensate contribution ${\cal L}^{\langle \alpha GG\rangle}_k(\tau)$ is less than 1/3 of perturbative contributions ${\cal L}^{\rm pert}_k(\tau)$. The convergence criterion \eqref{B_k} is independent of the anomalous dimension factors, and the resulting bound on $\tau$ will be compared for the LO and NLO perturbative contributions. 

The relative importance of the $\langle\alpha_s GG\rangle$ gluon condensate contributions in comparison with the other condensates in  \eqref{eq:rho_QCD} can be demonstrated via an extension of \eqref{B_k}  
\begin{equation}
    \tilde B_k^{\langle \bar q q\rangle}=\frac{{\cal L}^{\langle \bar q q\rangle}_k(\tau)}{{\cal L}^{\rm pert}_k(\tau)}\;\,,~
   \tilde B_k^{\langle \bar q q\rangle\langle \alpha GG\rangle}=\frac{{\cal L}^{\langle \bar q q\rangle\langle \alpha GG\rangle}_k(\tau)}{{\cal L}^{\rm pert}_k(\tau)}\,,
    \label{tilde_B_k}
\end{equation}
and thus $\tilde B_k\ll B_k $ would indicate dominance of the $\langle\alpha_s GG\rangle$ condensate contributions.

Establishing a lower bound on $\tau$ is often achieved via the ``pole contribution" criterion. Unfortunately this criterion is open to much broader interpretation in cases like the multi-quark scenario being considered because there is no guidance from explicit spectral function experimental data. However, the spectral function in \eqref{had_rho} satisfies $\rho^{\rm res}>0$, providing an integration measure $[d\mu]=\rho^{\rm res}(t)\,dt$, implying that the hadronic side of \eqref{LSR_final} must satisfy standard integral inequalities (e.g., Schwarz and H\"older inequalities). Thus the QCD quantity ${\cal R}_k\left(\tau,s_0\right)$ in \eqref{LSR_final} must satisfy the same (integral) inequalities to be consistent with an integrated positive hadronic spectral function. The following sum-rule H\"older inequality \cite{Benmerrouche:1995qa} will be employed
\begin{equation}
    \frac{\mathcal{R}_{k}\bigl(\tau+[1-\omega]\,\delta\tau,s_{0} \bigr)}{\mathcal{R}_{k}^{\omega}\bigl(\tau,s_{0}\bigr) \mathcal{R}_{k}^{1-\omega}\bigl(\tau+\delta\tau,s_{0}\bigr)}\leq 1,\,
    \quad 0\leq \omega\leq 1,
    \;\text{and } \;k\ge0\,,
    \label{eq:holderIneq1}
\end{equation}
where $\delta\tau=0.05 \,\rm{GeV}^{-2}$ will be used as in previous QCD applications 
(see e.g., Refs.~\cite{Benmerrouche:1995qa,Steele:1998ry,Shi:1999hm}) so that the inequality becomes local in the $\tau$, $s_0$ parameter space. 
The H\"older inequality constraint \eqref{eq:holderIneq1} is formed from a dimensionless ratio of Laplace sum-rules and hence anomalous dimension factors will be canceled. Similarly, a constraint based on the Schwarz inequality \cite{Kleiv:2013dta} will be used 
\begin{equation}
    \frac{\mathcal{R}_{k}(\tau,s_{0})/\mathcal{R}_{k-1}(\tau,s_{0})}{\mathcal{R}_{k-1}(\tau,s_{0})/\mathcal{R}_{k-2}(\tau,s_{0})}\ge 1,\, 
    \quad k\ge 2.
    \label{eq:holderIneq2}
\end{equation}
The Schwarz inequality constraint \eqref{eq:holderIneq2} is also formed from sum-rule ratios which therefore cancels the anomalous dimension factors.

Implementation of the Borel window criteria requires the dimension-four quark and gluon QCD condensates appearing in \eqref{eq:rho_QCD}. For the quark condensate, the PCAC relation \cite{Gell-Mann:1968hlm} is used 
\begin{equation}
    \left(m_u+m_d \right)\langle \bar q q\rangle=-f_\pi^2m_\pi^2\,,
\end{equation}
where in our conventions $f_\pi=130/\sqrt{2}\,{\rm MeV}$. For the gluon condensate, the determination from charmonium systems \cite{Narison:2011rn} will be employed
\begin{equation}
    \langle \alpha_s GG\rangle=\langle \alpha_s G^a_{\mu\nu}G^a_{\mu\nu}\rangle=\left( 0.075\pm 0.02
    \right)\,{\rm GeV^4}\,.
    \label{gluon_condensate_value}
\end{equation}
Ref.~\cite{Chen:2007xr} has used vacuum saturation to obtain the $\langle \bar q q\rangle\,\langle \alpha_s GG\rangle$ term in \eqref{eq:rho_QCD} and as demonstrated below, this term is small enough that the potentially large vacuum-saturation uncertainty has a negligible numerical effect.

\begin{figure}[htb]
    \centering
    \begin{minipage}{.45\textwidth}
        \centering
        \includegraphics[scale=0.78]{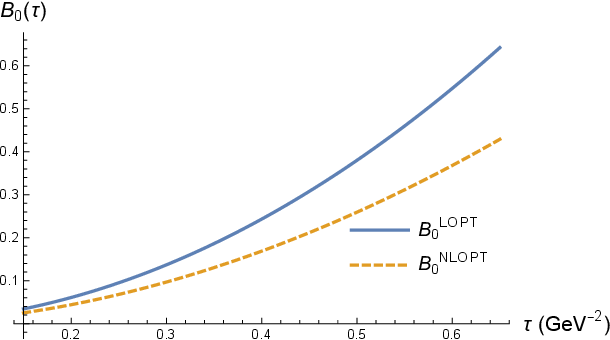}
    \end{minipage}
    \begin{minipage}{.45\textwidth}
        \centering
        \includegraphics[scale=0.78]{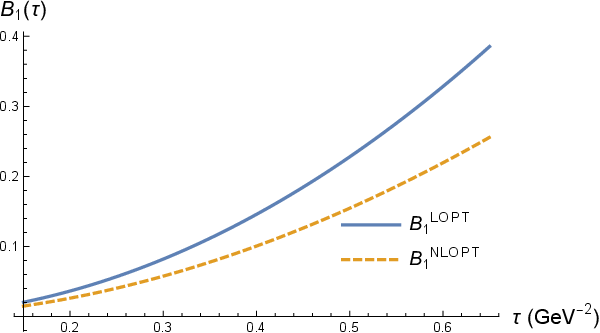}
    \end{minipage}
    \caption{Ratio $B_k$ [see Eq.~\eqref{B_k}] of the gluon condensate $\langle \alpha_s GG\rangle$ contributions to the LO (blue solid curves) and NLO (orange dashed curves) perturbative terms for $k=0,1$. Note that the NLO curves correspond to the total perturbative contributions up to NLO. }
   \label{fig:B_k}
\end{figure}

\begin{figure}[htb]
    \centering
    \begin{minipage}{.45\textwidth}
        \centering
        \includegraphics[scale=0.78]{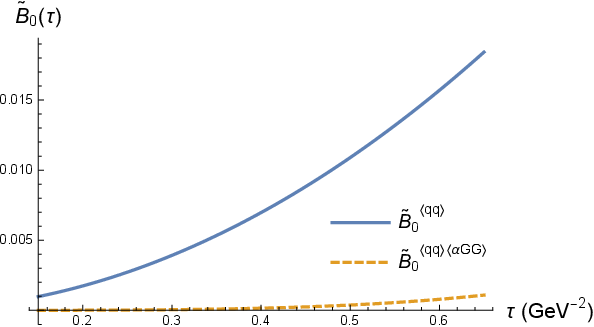}
    \end{minipage}
    \begin{minipage}{.45\textwidth}
        \centering
        \includegraphics[scale=0.78]{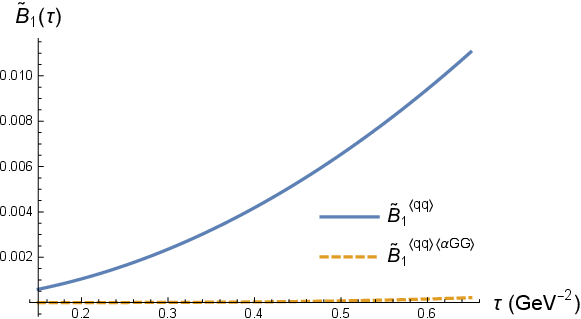}
    \end{minipage}
    \caption{Ratio $\tilde B_k$ [see Eq.~\eqref{tilde_B_k}] of the quark condensate $\langle \bar q q\rangle$ contributions to LO perturbation theory (blue solid curves) and for the mixed-condensate $\langle \bar q q\rangle\langle \alpha_s GG\rangle$ contributions (orange dashed curves) $k=0,1$. Note that in comparison to Fig.~\ref{fig:B_k}, $\tilde B_k$ is orders of magnitude smaller than $B_k$. }
    \label{fig:B_k_2}
\end{figure}

The upper bound on $\tau$ for the Borel window can now be examined using the convergence criterion for the ratio of non-perturbative (QCD condensate) and perturbative contributions to ${\cal L}_k(\tau)$. As shown in Figs.~\ref{fig:B_k} and \ref{fig:B_k_2}, the $\langle\alpha_s GG\rangle$ gluon condensate provides the dominant non-perturbative contribution ({\it i.e.,} $\tilde B_k\ll B_k$), and NLO effects increase the upper bound on $\tau_{\rm max}$ (see Table \ref{tab:GluonRatio}). This leads to our first important conclusion: NLO effects lead to a substantial increase of the $\tau$ upper bound for the Borel window, enhancing the reliability of the QCD Laplace sum-rule analysis. 

\begin{table}[htb]
    \centering
    \begin{tabular}{l|r|r}
      $k$ & $B^{\langle \alpha GG\rangle/\rm{LO\, PT}}_{k}$ & $B^{\langle \alpha GG\rangle/\rm{NLO\, PT}}_{k}$\\\hline\hline
        0 &  $\tau\leq 0.47$ GeV$^{-2}$ & $\tau\leq 0.57$ GeV$^{-2}$ \\
        1 & $\tau\leq 0.61$ GeV$^{-2}$ & $\tau\leq 0.75$ GeV$^{-2}$ \\\hline
    \end{tabular}
    \caption{Results of Eqs.~\eqref{B_k} for $k=0,1$, showing the increase of the Borel window upper bound when adding the NLO perturbative correction to the spectral function. The notation LO PT indicates that LO perturbative contributions are included, and NLO PT indicates perturbative contributions up to NLO (\textit{i.e}., LO plus NLO) are included. }
    \label{tab:GluonRatio}
\end{table}

In previous QCD sum-rule applications, the H\"older inequalities \eqref{eq:holderIneq1} often resulted in a lower bound on $s_0$ in addition to $\tau$ bounds for the Borel window \cite{Benmerrouche:1995qa,Steele:1998ry,Shi:1999hm}. Analysis of the inequalities \eqref{eq:holderIneq1} and \eqref{eq:holderIneq2} result in a lower bound $s_0\gtrsim 0.33\,{\rm GeV^2}$ at both LO and NLO, with an associated lower bound $\tau\gtrsim 0.2\,{\rm GeV^{-2}}$. The bounds are slightly weaker for the $k=1$ case than for $k=0$, so the $k=0$ bounds are adopted to ensure that both the $k=0$ and $k=1$ sum-rules satisfy the inequalities. Combining the H\"older inequality bounds with the convergence criteria constraints result in our final determination of the Borel window given in Table~\ref{borel_window_tab}.
 
\begin{table}[htb]
    \centering
    \begin{tabular}{l|c}
        Variable & Range\\\hline\hline
        $s_{0}$ & $\ge 0.33\,{\rm GeV^2}$\\
        $\tau$  &  $0.2\,\textrm{--} \,0.57\, {\rm GeV}^{-2}$
        \\
        \hline
    \end{tabular}
    \caption{Constraints on $s_0$ and $\tau$ defining the Borel-window working region for the QCD Laplace sum-rule analysis. The upper bound on $\tau$ emerges from the NLO $k=0$ entry of Table~\ref{tab:GluonRatio} and the lower bounds on $s_0$ and $\tau$ emerge from the $k=0$ inequalities \eqref{eq:holderIneq1} and \eqref{eq:holderIneq2}. }
    \label{borel_window_tab}
\end{table}
 
\begin{figure}[htb]
    \centering
    \includegraphics{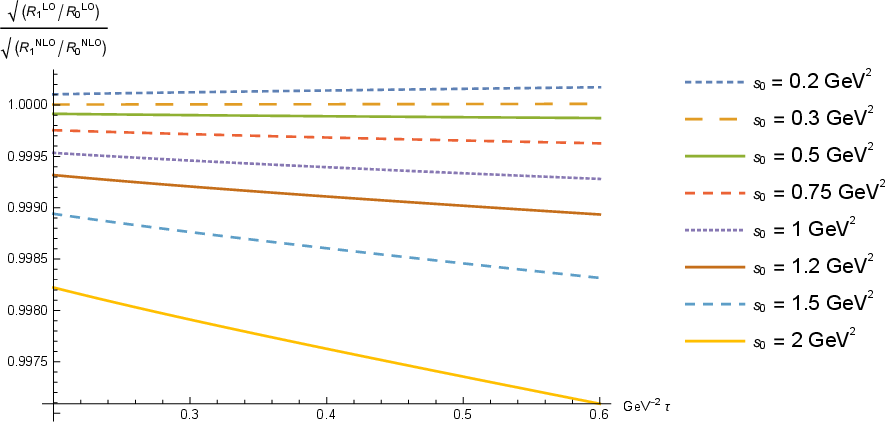}
    \caption{The double sum-rule ratio ${\cal R}_1\left(\tau,s_0\right)/{\cal R}_0\left(\tau,s_0\right)$ evaluated with inclusion of LO and NLO perturbative contributions is shown as a function of $\tau$ within the Borel window for selected $s_0$. Note that the ratio ${\cal R}_1\left(\tau,s_0\right)/{\cal R}_0\left(\tau,s_0\right)$ is independent of the anomalous dimension. }
    \label{sr_ratio_fig}
\end{figure}

Now that a Borel window working region for has been obtained, the effect of NLO corrections can be explored in more detail. Although the NLO perturbative effects are demonstrably large as shown in Fig.~\ref{Lk_NLO_fig}, the sum-rule ratio ${\cal R}_1\left(\tau,s_0\right)/{\cal R}_0\left(\tau,s_0\right)$, which cancels the anomalous dimension factors, is much less sensitive to NLO corrections. In Fig.~\ref{sr_ratio_fig} the double ratio of the LO and NLO sum-rule quantity ${\cal R}_1\left(\tau,s_0\right)/{\cal R}_0\left(\tau,s_0\right)$ is shown within the Borel window of Table~\ref{borel_window_tab}, demonstrating a remarkable stability under inclusion of NLO corrections. Thus we reach a second important conclusion that a Laplace sum-rule analysis based upon the ratio ${\cal R}_1\left(\tau,s_0\right)/{\cal R}_0\left(\tau,s_0\right)$ will be remarkably robust under the large NLO effects within the individual sum-rules. In Section~\ref{prediction_section} the ratio ${\cal R}_1\left(\tau,s_0\right)/{\cal R}_0\left(\tau,s_0\right)$ will be used to study the predictions for resonance models of increasing complexity. 

\section{Laplace Sum-Rule Analyses of Resonance Models} 
\label{prediction_section}
\begin{figure}[htb]
    \centering
    \includegraphics{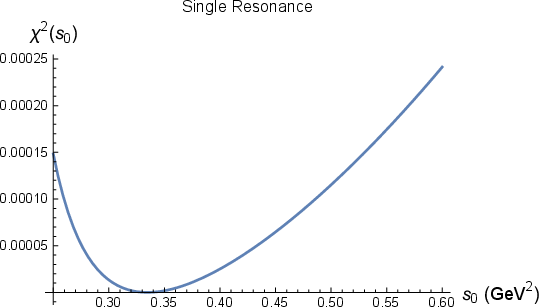}
    \caption{The single-resonance model residual $\chi_{\text{SR}}^{2}$ is shown as a function of the continuum threshold $s_0$ for the optimized $m_\sigma$ (defined as a function of $s_0$). }
    \label{sr_chi2_fig}
\end{figure}

The simplest resonance model is a single narrow resonance (SR)
\begin{equation}
\rho^{\rm SR}(t)=A_\sigma\delta\left(t-m_\sigma^2\right)
\label{sr_model}    
\end{equation}
where $A_\sigma>0$ parameterizes the coupling of the resonance to the tetraquark current \eqref{eq:current1} and the resonance mass is tentatively associated with the $\sigma$ meson (\textit{i.e.}, $f_0(500)$ in PDG listings \cite{ParticleDataGroup:2020ssz}). Using \eqref{sr_model} for $\rho^{\rm res}(t)$ in \eqref{LSR_final} gives
\begin{equation}
    \frac{\mathcal{R}_{1}(\tau,s_{0})}{\mathcal{R}_{0}(\tau,s_{0})} = m^{2}_{\sigma}\,,
    \label{eq:QCDSRSingle}
\end{equation}
where the formulation as a ratio of sum-rules has been chosen to cancel the anomalous dimension factors. The mass $m_\sigma$ and continuum $s_0$ can then be predicted by minimizing the residual sum of squares
\begin{equation}
        \chi_{\text{SR}}^{2}(s_{0})= \sum_{j} \Biggl( m_{\sigma}^{2} \frac{\mathcal{R}_{0}(\tau_{j},s_{0})}{\mathcal{R}_{1}(\tau_{j},s_{0})} -1\Biggr)^{2}\,,
    \label{eq:chi2Single}
\end{equation}
where the sum is over discrete, equally spaced $\tau$ values ($\tau_{j+1}-\tau_j=0.05\,{\rm GeV^{-2}}$) in the Borel window and there is an implicit lower bound constraint on $s_0$ (see Table~\ref{borel_window_tab}). The optimized $m_\sigma$ can be expressed as a function of $s_0$, resulting in the single-variable residual shown in Fig.~\ref{sr_chi2_fig}. The single-resonance model predictions obtained by minimizing \eqref{eq:chi2Single} are given in Table~\ref{predictions_table}. Note that the residual sum of squares is defined by the Borel window, so the influence of changes in the Borel window (e.g., via higher-loop corrections) on the optimized predictions of $m_\sigma$ and $s_0$ has also been explored in Table~\ref{predictions_table}.

\begin{table}[htb]
\centering
    \begin{tabular}{l|ccrcc}
    Model  & $s_{0}$ (GeV$^{2}$) Window & $\tau$ (GeV$^{-2}$) Window & $m_{\sigma}$ (GeV) & $s_0^{\text{opt}}$ (GeV$^{2}$) & $r$  \\\hline\hline
    SR & $0.33$ 
    -- $1.3$ 
    & $0.2 - 0.57$ & $0.52$& 0.335 &-- \\
    DR & $0.98$ -- $1.3$ & $0.2$ -- $0.57$
    &$0.68$--$0.77$
    & 
   $1.04$--$1.12$
    & 
    $1.18$--$3.38$
    \\\hline
\end{tabular}
\caption{ 
Predictions for the optimized mass and continuum threshold in single narrow (SR) and double narrow (DR) resonance models. The windows for $s_0$ and $\tau$ follow from the inequality bounds of Table~\ref{borel_window_tab} and for the DR case there is an implicit constraint that $\sqrt{s_0}$ is larger than the $f_0(980)$ mass. The upper bound on the $s_0$ window is a constraint on the numerical search for a minimum in the residual sum of squares. The range of values given for the predicted values of $\{m_\sigma,s_0^{\rm opt},r\}$ reflects the theoretical uncertainties described in the text. With the inclusion of resonance width effects,  Table entries  for $m_\sigma$ and $r$ are interpreted as the effective mass and coupling ratio $m^{\rm eff}_\sigma$ and $r^{\rm eff}$ as used in Eqs.~\eqref{SW_eff} and \eqref{AW_eff}.
}
\label{predictions_table}
\end{table}

Although the mass prediction in the SR single-resonance model of Table~\ref{predictions_table} is in good agreement with the observed $f_0(500)$ \cite{ParticleDataGroup:2020ssz}, the optimized $s_0$ is uncomfortably close to the lower bound permitted by the H\"older inequalities (Table~\ref{borel_window_tab}). Furthermore, the gap between the resonance and continuum is very small, particularly if the model is extended to include resonance width effects. These challenges in the single-resonance model can be addressed by including a second resonance, shifting part of the spectral function strength from the continuum to the heavier state and thereby raising the value of $s_0$. 

\begin{figure}[htbp]
    \centering
    \includegraphics[scale=0.96]{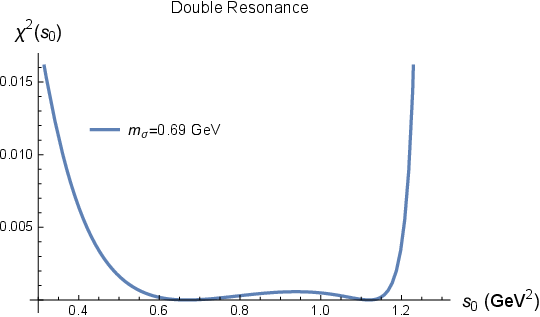}
     \includegraphics[scale=0.96]{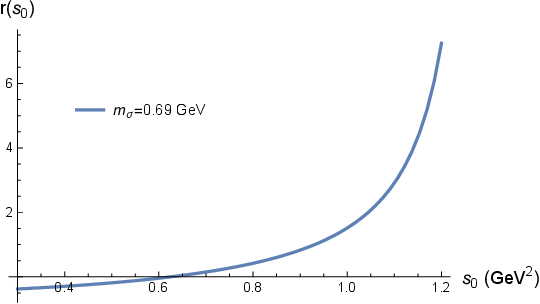}
    \caption{The coupling ratio $r$ \eqref{dr_parameters} and the residual sum of squares \eqref{eq:chi2Double} is shown as a function of $s_0$ for central values of input parameters and the optimized $m_\sigma=0.69\,{\rm GeV}$ central value from Table~\ref{predictions_table}. }
    \label{dr_plots}
\end{figure}

The double narrow resonance model (DR) is given by
\begin{equation}
\rho^{\rm DR}(t)=A_\sigma\delta\left(t-m_\sigma^2\right)+A_{f_0}\delta\left(t-m^2_{f_0}\right)\,,
\label{dr_model}    
\end{equation}
where $A_{f_0}>0$ parameterizes the coupling of the heavier state to the tetraquark current \eqref{eq:current1}. Using \eqref{dr_model} for $\rho^{\rm res}(t)$ in \eqref{LSR_final} gives
\begin{equation}
    \frac{\mathcal{R}_{1}(\tau,s_{0})}{\mathcal{R}_{0}(\tau,s_{0})} =  \frac{A_{\sigma}m_{\sigma}^{2} \,e^{-m_{\sigma}^{2}\tau} +A_{f_{0}}m_{f_{0}}^{2}\,e^{-m_{f_{0}}^{2}\tau}}{A_{\sigma} \,e^{-m_{\sigma}^{2}\tau} +A_{f_{0}}\,e^{-m_{f_{0}}^{2}\tau} }\,.
    \label{eq:QCDSRDouble}
\end{equation}
Although there will be an exponential suppression of the heavier state in ${\cal R}_0$, the large mass hierarchy $m^2_{f_0}/m^2_\sigma\approx 4$ provides a compensating enhancement of the heavier state's signal in ${\cal R}_1$, and thus it is meaningful to explore the double resonance scenario. Because the $f_0(980)$ is well-established, the central value $m_{f_0}=0.98\,{\rm GeV}$ will be used as an   input parameter and minimization of the following residual sum of squares will be used to determine the remaining parameters of the double resonance model
\begin{gather}
    \chi_{\text{DR}}^{2}(s_{0})= \sum_{j} \Biggl( m_{\sigma}^{2}\frac{\mathcal{R}_{0}(\tau_{j},s_{0})}{\mathcal{R}_{1}(\tau_{j},s_{0})} - 1 + r \,m_{f_{0}}^{2}e^{-\Delta m^{2}\tau_{j}}\frac{\mathcal{R}_{0}(\tau_{j},s_{0})}{\mathcal{R}_{1}(\tau_{j},s_{0})} - r e^{-\Delta m^{2}\tau_{j}} \Biggr)^{2}\,,
    \label{eq:chi2Double}
    \\
    r=\frac{A_{f_0}}{A_\sigma}\,,~\Delta m^2=m^2_{f_0}-m^2_\sigma\,,
\label{dr_parameters}
\end{gather}
where the sum is over discrete, equally spaced $\tau$ values ($\tau_{j+1}-\tau_j=0.05\,{\rm GeV^{-2}}$) in the Borel window (see Table~\ref{borel_window_tab}) and the formulation has been chosen to contain sum-rule ratios to cancel the anomalous dimensions. With the central value input $m_{f_0}=0.98\,{\rm GeV}$, the optimized value of $r$ can be determined as a function of $s_0$ and $m_\sigma$, and a numerical minimization of the residual sum of squares is performed in the resulting two-dimensional $\{m_\sigma,s_0\}$ parameter space. 
Theoretical uncertainties in the resulting predicted values of $\{m_\sigma,s_0^{\rm{opt}},r\}$ are modeled by
variations of  $\langle \alpha_s GG\rangle$ in \eqref{gluon_condensate_value} and associated changes in the Borel window, variations of  $f_0(980)$ mass within the PDG range \cite{ParticleDataGroup:2020ssz}, and  independent variations of $\pm 0.05\,{\rm GeV^{-2}}$ in the Borel window. The results shown in Table \ref{predictions_table} have a much better separation of the continuum from the $f_0(980)$ (the heavier resonance), while still predicting a $\sigma$ mass consistent with the PDG values \cite{ParticleDataGroup:2020ssz}. Fig.~\ref{dr_plots} shows the $s_0$ dependence of the coupling ratio $r$ \eqref{dr_parameters} and the residual sum of squares \eqref{eq:chi2Double} for central values of input parameters and the optimized central value $m_\sigma=0.69\,{\rm GeV}$. The local minimum of $\chi_{\text{DR}}^{2}$ at $s_0\approx 1.1\,{\rm GeV^2}$ is deeper than the local minimum at $s_0\approx 0.65\,{\rm GeV^2}$, which identifies $s_0\approx 1.1\,{\rm GeV^2}$ as the global minimum. However, the false minimum $s_0\approx 0.65\,{\rm GeV^2}$ is qualitatively consistent with the $s_0$ value found in the single resonance analysis because $r\approx 0$ in this $s_0$ region, decoupling the $f_0(980)$. The  Table~\ref{predictions_table} result $r>1$ implies that the $f_0(980)$ coupling to the (non-strange) tetraquark current \eqref{eq:current1} is enhanced compared to the $\sigma$ [\textit{i.e.}, $f_0(500)$], compensating for the $\exp{\left(-\Delta m^2\tau \right)}$ exponential suppression of the $f_0(980)$ contribution to \eqref{eq:chi2Double}. A similar $f_0(980)$ coupling enhancement has also been found in chiral Lagrangian analyses which find a larger non-strange tetraquark component for the $f_0(980)$ compared to the $f_0(500)$ \cite{Fariborz:2003uj,Fariborz:2009cq}. 

\begin{figure}[htb]
    \centering
    \includegraphics[scale=1.2]{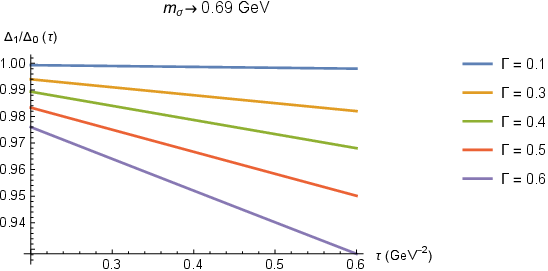}
    \caption{
    The quantity $\Delta_1/\Delta_0$ is shown as a function of $\tau$ within the Borel window for $m_\sigma=0.69\,{\rm GeV}$ (see Table~\ref{predictions_table}) and for selected values of $\Gamma_\sigma$.
   }
    \label{Delta_fig}
\end{figure}

The $f_0(500)$ is a notoriously broad state \cite{ParticleDataGroup:2020ssz}, so the final refinement of our analysis is to extend the two-resonance model to include width effects for the $f_0(500)$ (the $f_0(980)$ is quite narrow, so its width effects are numerically small in the Laplace sum-rules). The simplest symmetric width (SW) model symmetrically distributes the resonance strength over the total width parameterized by $\Gamma_\sigma$ \cite{Elias:1998bq}
\begin{equation}
\rho^{\rm SW}(t)=\frac{A_\sigma}{2m_\sigma \Gamma_\sigma}
\left[
\theta\left(t-m_\sigma^2+m_\sigma\Gamma_\sigma\right) 
-\theta\left(t-m_\sigma^2-m_\sigma\Gamma_\sigma\right)
\right]\,,
\label{square_pulse_rho}
\end{equation}
where the normalization has been chosen so that \eqref{square_pulse_rho} and \eqref{sr_model} have the same integrated resonance strength $A_\sigma$. The resulting SW modification of the single-resonance model \eqref{eq:QCDSRSingle} is given by \cite{Elias:1998bq}
\begin{gather}
    \frac{\mathcal{R}_{1}(\tau,s_{0})}{\mathcal{R}_{0}(\tau,s_{0})} = m^{2}_{\sigma}\frac{\Delta_1\left(m_{\sigma},\Gamma,\tau\right)}{\Delta_0\left(m_{\sigma},\Gamma,\tau\right)}
    \,,
    \label{SW_single}
    \\[5pt]
    \Delta_{0}\left(m_{\sigma},\Gamma,\tau\right) = \frac{\sinh{\left(m_{\sigma}\Gamma_{\sigma}\tau\right)}}{m_{\sigma}\Gamma_{\sigma}\tau},
    \\[5pt]
    \Delta_{1}\left(m_{\sigma},\Gamma,\tau\right) = \left(1+ \frac{1}{m_{\sigma}^{2}\tau}\right)
    \Delta_{0}\left(m_{\sigma},\Gamma,\tau\right)- \frac{\cosh{\left(m_{\sigma}\Gamma_{\sigma}\tau\right)}}{m_{\sigma}^{2}\tau}\,,
\end{gather}
and similarly modifies the double-resonance model \eqref{eq:QCDSRDouble} to
\begin{equation}
    \frac{\mathcal{R}_{1}(\tau,s_{0})}{\mathcal{R}_{0}(\tau,s_{0})} =  \frac{A_{\sigma}\Delta_1 m_{\sigma}^{2} \,e^{-m_{\sigma}^{2}\tau} +A_{f_{0}}m_{f_{0}}^{2}\,e^{-m_{f_{0}}^{2}\tau}}{A_{\sigma}\Delta_0 \,e^{-m_{\sigma}^{2}\tau} +A_{f_{0}}\,e^{-m_{f_{0}}^{2}\tau} }=
    \frac{\Delta_0 A_{\sigma}\left(m_{\sigma}^{2}\frac{\Delta_1}{\Delta_0}\right) \,e^{-m_{\sigma}^{2}\tau} +A_{f_{0}}m_{f_{0}}^{2}\,e^{-m_{f_{0}}^{2}\tau}}{\Delta_0 A_{\sigma} \,e^{-m_{\sigma}^{2}\tau} +A_{f_{0}}\,e^{-m_{f_{0}}^{2}\tau} }
    \,.
    \label{SW_double}
\end{equation}
For $\Gamma_\sigma<0.6\,{\rm GeV}$ and for the $m_\sigma$ parameters of Table \ref{predictions_table}, $\Delta_1$ and $\Delta_0$ are nearly constant in the $\tau$ range of the Borel window (less than 5\% variation) and hence the fitting procedures developed above are modified by width effects to predicting an effective sigma mass $m_\sigma^{\rm eff}$ and effective coupling ratio $r^{\rm eff}$ related to the physical mass $m_\sigma$ and coupling ratio $r$ by
\begin{equation}
\left(m^{\rm eff}_\sigma\right)^2=m_\sigma^2\frac{\Delta_1}{\Delta_0}
\,,~
r^{\rm eff}=\frac{r}{\Delta_0}\,.
\label{SW_eff}
\end{equation}
Fig.~\ref{Delta_fig} shows that for $\Gamma_\sigma<0.6\,{\rm GeV}$ the central value effective mass $m_\sigma^{\rm eff}=0.69\,{\rm GeV}$ is enhanced by less than 5\%, and using the corresponding value of $\Delta_0$, the effective coupling ratio $r^{\rm eff} \approx 3$ is suppressed by less than 1\%. Because any symmetric resonance shape (e.g., Breit-Wigner) can be approximated numerically within the Laplace sum-rules as a Riemann sum (linear combinations of \eqref{square_pulse_rho} with different $A_\sigma$ and $\Gamma_\sigma$) \cite{Elias:1998bq}, width effects from symmetric resonance shapes will have similarly small numerical effects.

\begin{figure}[htb]
    \centering
    \includegraphics[scale=1.5]{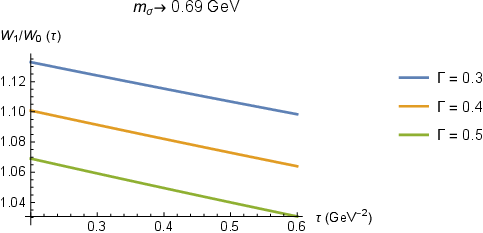}
    \caption{The quantity $W_1/W_0$ is shown as a function of $\tau$ within the Borel window for $m_\sigma=0.69\,{\rm GeV}$ (see 
    Table~\ref{predictions_table}) and for selected values of $\Gamma_\sigma$.}
    \label{fig:Assym_shape_DR}
\end{figure}

For an asymmetric $\sigma$ resonance shape, chiral Lagrangian methods for $\pi\pi$ scattering \cite{Sannino:1995ik,Harada:1995dc,Harada:1996wr,Black:1999dx} can be used to obtain the following asymmetric (AW) resonance model in the $m_\pi=0$ limit \cite{Shi:1999hm}
\begin{equation}
   \rho^{\rm AW}(s) = \frac{A_\sigma}{N m^2_{\sigma} }\frac{s^{2}}{\left[(s-m_{\sigma}^{2})^{2} + \frac{\Gamma^{2}_{\sigma}}{m_{\sigma}^{6}}s^{4}\right]},
   \label{rho_AW}
\end{equation}
where $N$ is a normalization factor to ensure that \eqref{rho_AW} and \eqref{sr_model} have the same integrated resonance strength. The AW modifications to the single and double resonance model analogous to \eqref{SW_single} and \eqref{SW_double} are 
\begin{gather}
    \frac{\mathcal{R}_{1}(\tau,s_{0})}{\mathcal{R}_{0}(\tau,s_{0})} = m^{2}_{\sigma}\frac{W_1\left(m_{\sigma},\Gamma,\tau,s_0\right)}{W_0\left(m_{\sigma},\Gamma,\tau,s_0\right)}
    \,,
    \\[5pt]
    W_{k} \left(m_{\sigma},\Gamma_{\sigma},\tau,s_{0}\right) =\frac{1}{N} \int_{-1}^{\frac{s_{0}-m_{\sigma}^{2}}{m_{\sigma}^{2}}} \,d\zeta\,e^{-m_{\sigma}^{2}\zeta\tau}\,\frac{(\zeta+1)^{2+k}}{\zeta^{2}+\frac{\Gamma_{\sigma}^{2}}{m_{\sigma}^{2}}(\zeta+1)^{4}}
    \,,
    \\
    N=\int_0^\infty d\zeta \frac{\zeta^2}{\left(\zeta-1\right)^2+\frac{\Gamma_\sigma^2}{m_\sigma^2}\zeta^4}
    \,,
    \\
     \frac{\mathcal{R}_{1}(\tau,s_{0})}{\mathcal{R}_{0}(\tau,s_{0})} =  \frac{A_{\sigma}W_1 m_{\sigma}^{2} \,e^{-m_{\sigma}^{2}\tau} +A_{f_{0}}m_{f_{0}}^{2}\,e^{-m_{f_{0}}^{2}\tau}}{A_{\sigma}W_0 \,e^{-m_{\sigma}^{2}\tau} +A_{f_{0}}\,e^{-m_{f_{0}}^{2}\tau} }=
    \frac{W_0 A_{\sigma}\left(m_{\sigma}^{2}\frac{W_1}{W_0}\right) \,e^{-m_{\sigma}^{2}\tau} +A_{f_{0}}m_{f_{0}}^{2}\,e^{-m_{f_{0}}^{2}\tau}}{W_0 A_{\sigma} \,e^{-m_{\sigma}^{2}\tau} +A_{f_{0}}\,e^{-m_{f_{0}}^{2}\tau} }
    \,.
\end{gather}
As in the SW case, for the Table \ref{predictions_table} parameters and the range $0.3\,{\rm GeV}<\Gamma_\sigma<0.5\,{\rm GeV}$ characteristic of PDG values \cite{ParticleDataGroup:2020ssz}, the quantities $W_1/W_0$ and and $W_0$ are nearly constant in the $\tau$ Borel window, and hence the fitting procedures developed above are again modified by width effects to predicting an effective sigma mass $m_\sigma^{\rm eff}$ and effective coupling ratio $r^{\rm eff}$ related to the physical mass $m_\sigma$ and coupling ratio $r$ by
\begin{equation}
\left(m^{\rm eff}_\sigma\right)^2=m_\sigma^2\frac{W_1}{W_0}
\,,~
r^{\rm eff}=\frac{r}{W_0}\,.
\label{AW_eff}
\end{equation}
Fig.~\ref{fig:Assym_shape_DR} shows that for $0.3\,{\rm GeV}<\Gamma_\sigma<0.5\,{\rm GeV}$ the central value effective mass $m_\sigma^{\rm eff}=0.69\,{\rm GeV}$ is suppressed by approximately 5\%, and using the corresponding value of $W_0$, the effective coupling ratio $r^{\rm eff}\approx 3$ is enhanced by approximately 30\%. The AW model suppression of the predicted value of $m_\sigma$ compared with SW models is consistent with the trend found in Ref.~\cite{Shi:1999hm}. The narrow-width mass predictions of Table \ref{predictions_table} are thus surprisingly robust under resonance width effects, while the relative resonance strength of the $f_0(980)$ should be interpreted as an underestimate by $\sim 30\%$ associated with different $\sigma$ width models.

\section{Conclusions}
\label{conclusions_section}

QCD sum-rule analyses of tetraquark systems have primarily been performed to LO in perturbation theory because of the technical challenges (e.g., loop topologies with massive quarks, complicated operator mixing under renormalization) of calculating NLO corrections. Given the multitude of sum-rule analyses in tetraquark systems (see e.g., Refs.~\cite{Liu:2019zoy,Albuquerque:2018jkn}) it is important to determine the impact of NLO perturbative corrections on the Laplace sum-rule predictions of hadronic properties.

In this paper, we have used the analytic results for NLO perturbative corrections to the scalar-isoscalar correlator \cite{Groote:2014pva} to perform a detailed case study comparing LO and NLO Laplace sum-rule analyses. As shown in Fig.~\ref{Lk_NLO_fig}, the NLO corrections are large in the individual Laplace sum-rules, indicative of slow perturbative convergence of the underlying QCD correlator. Furthermore, the NLO contributions can be used to extract the anomalous dimension for the QCD spectral function given in \eqref{gamma_rho}, and because this anomalous dimension could contain large unknown higher-order corrections, our methodology has focused on sum-rule ratios to remove the anomalous dimension factor.

Although the large numerical effect of the NLO corrections seems problematic, Table~\ref{tab:GluonRatio} shows that the $\tau$ Borel window is expanded by the NLO corrections, providing a stronger foundation for a sum-rule analysis. Furthermore, within this Borel window, Fig.~\ref{sr_ratio_fig} shows that ratios of sum-rules are much less sensitive to the large NLO perturbative corrections, and have the added advantage of being independent of the anomalous dimension. Thus despite the superficially large NLO perturbative corrections, sum-rule predictions obtained from ratios of the light tetraquark scalar-isoscalar Laplace sum-rules are surprisingly robust. 

The NLO tetraquark Laplace sum-rules were then used to obtain predictions for the lightest scalar-isoscalar $\sigma$ state in a variety of models, including multiple states and $\sigma$ width effects. The Table \ref{predictions_table} predictions are in good agreement with the $f_0(500)$ PDG mass values \cite{ParticleDataGroup:2020ssz}, and indicate a relative tetraquark component of the $f_0(980)$ consistent with chiral Lagrangian analyses \cite{Fariborz:2003uj,Fariborz:2009cq}.  

In conclusion, our detailed case study of NLO perturbative effects in the light tetraquark scalar-isoscalar channel provides grounds for cautious optimism that LO QCD sum-rule studies of other light-quark tetraquark systems may be unexpectedly robust under large NLO corrections, provided that analysis methodologies focused on sum-rule ratios are employed. It is not clear whether sum-rule ratio methods for heavy-quark multiquark systems will be similarly robust in situations with large NLO corrections.
However, if the patterns found for the scalar-isoscalar channel case study persist, NLO effects in  tetraquark sum-rules would modify LO sum-rule analyses in two key aspects: through perturbatively-controlled changes to sum-rule ratios, and via significant changes in the Borel window. The combination of these two aspects could improve stability of LO QCD sum-rule analyses (including channels that fail to stabilize at LO) and lead to interesting shifts in the LO predictions of hadronic properties.

\section*{Acknowledgments}
TGS is grateful for research funding from the Natural Sciences and Engineering Research Council of Canada (NSERC). We are grateful to S.~Groote for helpful discussions.

\clearpage
\end{document}